# The role of consumer networks in firms' multi-characteristics competition and market share inequality

Antonios Garas[a]

[a] *Chair of Systems Design ETH Zurich, WEV G 202.1 Weinbergstrasse 56/58, CH-8092 Zürich;* agaras@ethz.ch;

Athanasios Lapatinas[1, b, c]

[b] *European Commission, DG Joint Research Centre, Unit I.1. Modelling, Indicators and Impact Evaluation, Competence Centre on Microeconomic Evaluation, Via E. Fermi 2749, TP 361, Ispra (VA), I-21027, Italy;* athanasios.lapatinas@ec.europa.eu

[c] *Department of Economics, University of Ioannina, Greece;* alapatin@cc.uoi.gr

**Abstract**

We develop a location analysis spatial model of firms' competition in multi-characteristics space, where consumers' opinions about the firms' products are distributed on multilayered networks. Firms do not compete on price but only on location upon the products' multi-characteristics space, and they aim to attract the maximum number of consumers. Boundedly rational consumers have distinct ideal points/tastes over the possible available firm locations but, crucially, they are affected by the opinions of their neighbors. Proposing a dynamic agent-based analysis on firms' location choice we characterize multi-dimensional product differentiation competition as adaptive learning by firms' managers and we argue that such a complex systems approach advances the analysis in alternative ways, beyond game-theoretic calculations.

---

[1] Corresponding author

# 1. Introduction

Product characteristics are typically considered as given when economists study firms' strategies and behaviour. But firms in industries with product differentiation actually choose the features of their products based on consumers' preferences. The role and effects of the demand side in product differentiation processes have been neglected by the relevant literature (Coombs et al., 2001; Mueller et al., 2015) and only recently have started to receive attention (Andersen, 2007). The importance of consumer behaviour in shaping markets' properties is oversighted and in most cases the simplistic perspective of perfect rational homogeneous agents is implicitly adopted in the literature (Valente, 2012). Among the works highlighting the relevance of demand for the emergence of new products, Witt (2001), Saviotti (2001), Malerba et al. (2007), Windrum et al. (2009), Nelson and Consoli (2010), Valente (2012), Markey-Towler and Foster (2013), Mueller et al. (2015), Markey-Towler (2016) and Schlaile et al. (2017) can be included. However, no attention has been paid to advancing a generalized evolutionary model of demand for the effect of consumers' social networks on their consumption decisions and, in turn, on firms' process of developing new characteristics for their products.

The aim of this paper is to address this gap in the literature by suggesting a way to understand the relevance of consumers' social networks in shaping demand behaviour in such markets, which in turn frame the process of product-embodied innovations. We do so using complex networks, which mathematically are described by graph objects. We develop a stylized computational model to examine the issue of product differentiation in a multi-dimensional space with consumers' choices emerging from consumers' opinion-based multilayered networks. The model that we propose is designed to analyze and illustrate potential effects of consumers' heterogeneity, particularly regarding their preferences, their bounded rationality and the role of their social networks on the development of new product characteristics. We ask two questions: First, how consumers' social networks affect their decisions on which products to buy and how this decision leads to market-share inequality at the firm level? Second, how firms respond to these decisions by developing new characteristics for their products defined over a multidimensional characteristic space?

Assuming that bounded rational consumers with heterogeneous preferences can have an active role in the product-embodied innovation process, and studying the influence of a change in their purchasing opinion over time due to their participation on social networks calls for new modelling approaches, namely complex systems analysis and agent-based simulation techniques (Pyka and Grebel, 2006; Pyka and Fagiolo, 2007; Gilbert, 2008; Farmer and Foley, 2009; Tubaro, 2011; Schlaile et al., 2017; Muller, 2017). Agent-based simulation approaches focus on the rules and elements that constitute a system and the interactions of the players/agents within the system and have gained increasing momentum in many scientific disciplines (Schlaile et al., 2017; Mueller and Pyka, 2017; Vermeulen and Pyka, 2016; Namatame and Chen, 2016; Hamill and Gilbert, 2016; Boero et al., 2015; Wilensky and Rand, 2015; de Marchi and Page, 2014; Kiesling et al., 2012). In our case, through agent-based simulation we are able to account for the economic agents' heterogeneity and the related implications of their interactions by representing the economic system in a more realistic fashion overcoming the simplistic models limited to representative agents (Bonabeau, 2002; Macal and North, 2005, 2006, 2010, 2014; Page, 2011; Schalaile et al., 2017).

The agent-based simulation model described in this paper is based on the works by Laver, (2005), Laver and Sergenti (2011), Valente (2012), Markey-Towler and Foster (2013), Halu et al. (2013), Mueller et al. (2015), Schlaile et al. (2017), Muller (2017) and represents a first attempt to address the issue of neglected consumers' social networks in the process of developing new products. In particular, we focus on the effects of heterogeneous and boundedly rational consumers' social networks on firms' multi-dimensional product differentiation competition.[2] Our model captures only parts of the consumers' demand-side, namely 'consumption as voting' (Dickinson and Carsly, 2005; Shaw et al., 2006; Moraes et al., 2011) whereby consumers choose which firms/suppliers and new products/services to support.[3] In this sense, our model could also apply to political competition, as studied by Downs (1957) and many others after him. More specifically, it could be perfectly placed within the scope of agent-based models described in Laver and Sergenti (2011). However, in this work we will stick to the firm affairs' language.

---

[2] Illuminative for the reader might be the work by Markey-Towler (2016) who develops a very interesting general theoretical model of market dynamics in the field of evolutionary economics that elaborates the process of firms' competition for market-share on the basis of both prices and product characteristics.

[3] See also Schlaile et al. (2016) and Schlaile et al. (2017) for a thorough discussion on possible ways/domains that consumers take an active role in the process of developing new product characteristics.

In more detail, what we propose here is a multi-dimensional agent-based model of firms' locational choice in the product-characteristics space that describes a finite number of firms competing for customers.[4] This model is inspired by the New Consumer Theory literature dealing with the non-price aspects of consumer behaviour (Lancaster, 1966, 1971, 1975, 1979; Ironmonger, 1972; Ratchford, 1975; Earl, 1983, 1986) and has its roots in the behavioral economics literature having to do with boundedly rational consumers making decisions under the presence of multiple attributes (Selten, 1998). From consumers' perspective, we begin with a very basic idea. Consumers' decision on purchases cannot be different from most other decisions people make in their daily lives, in the sense that some process for acquiring information and evaluating it is necessary. The model also allows for heterogeneity of demand based on heterogeneous consumers having individual preferences for the particular characteristics of the products, like the colour, the size, the brand, the shape, their functionality etc.: consumers have preferences that place them on their ideal points in a multi-characteristics space.[5] For every firm we model the network of consumers' opinions about its product (i.e. for every firm there is a corresponding opinion network describing consumers' opinion about the firm's product: the nodes represent the consumers and a link between two nodes represents their exchange of opinion about the corresponding firm's product), on which (opinion network) a contagion dynamics can take place.[6] Consumers are represented on every network and can be active only in one of the networks (purchase of one product from one firm) at the end of a given time period (at the moment of the purchase). Each consumer has also the option not to purchase, and in that case she will be inactive in all networks. Before the end of each time period, and while the consumers think about the purchase, they are more likely to change opinions and examine different options. They have some form of product characteristics expectations at the beginning, which is reflected by their position in the product characteristics space, but their decision on which product to purchase is crucially affected by the opinion of their peers: at the time of the purchase, they tend to be active in the network where the majority of their peers are also active.[7] In addition, as they gain experience and acquire more information about the products, their opinion can change, but it stabilizes over time (see Hoeffler and Ariely, 1999; West et al., 1996). This "uncertainty reduction", reflecting both consumers' learning and changes in their environment, is captured by dynamics that slow down until the purchase moment when they become completely frozen. These dynamics are implemented in this work with the simulated annealing algorithm.[8] An illustration of a possible final configuration of the described system is shown in Figure 1.

---

[4] Other firm-level models of multidimensional product choice within the agent-based framework are, for example, Page and Tassier (2007), Babutsidze (2015). For models on product innovation see also Windrum and Birechenhall (1998), Chen and Chie (2005, 2007), Klette and Kortum (2004), Marengo and Valente (2010), Ciarli et al. (2010), Lorentz et al. (2016), Acemoglu et al. (2013), Georges (2015).
[5] Other models simulating a multidimensional characteristics space are, for example, Saviotti and Metcalfe (1984), Saviotti (1996), Gallouj and Weinstein (1997).
[6] See also Rauch and Casella (2001), Dutta and Jackson (2003), Earl and Potts (2004), Granovetter (2005).

At the end of the simulated annealing calculation we have the number of consumers that opt to purchase from each firm. This is affected by the average connectivities of the consumers' opinion networks. Hence, we observe "regions" in the average connectivity space where the active nodes belonging to some opinion networks with high average connectivity percolate the system, while nodes of the remaining opinion networks with lower average connectivities are concentrated in disconnected clusters indicating high market share inequality. Nevertheless, "regions" where the average connectivities of the consumers' opinion networks are comparable manifest a market that sustains low market share inequality (see also Halu et al., 2013).

To the best of our knowledge, this paper is the first study that incorporates the role of opinion exchange and processing on a multilayered network, in an attempt to shed light on how the market-share inequality and the competition in a multi-characteristics space are affected by the presence of consumers' interactions.[9] Taking into account these multiple layers is crucial, as the considerable interest in various multilayered systems demonstrates (Buldyrev et al., 2010; Vespignani, 2010; Parshani et al., 2011; Baxter et al., 2012; Bashan et al., 2012; Gomez et al., 2013; Radicchi and Arenas, 2013; De Domenico et al., 2013; Radicchi, 2014; Garas, 2016). On the other hand, to make significant advances in understanding consumers' decisions, we must device a method for studying this process. Such method would allow us to observe consumers' behaviour from up close, to dig below the surface and watch consumers as they try to exchange information about a myriad of alternative products while refining the overwhelmingly volume of information of their multiple characteristics. Behavioral decision theory guides the process-oriented, complex systems-framework we present in the next sections in an effort to develop a new set of measures for studying market proceedings. The results reported in the subsequent sections are intended to demonstrate that such a realistic complex systems approach provides a plausible basis for understanding market-share inequality and firms' competition in the product-characteristics space and can advance the analysis of these topics in interesting and alternative ways, beyond game-theoretic calculations.

---

[7] Two consumers are peers when there is a link between them, i.e. when they exchange opinions.
[8] An early application of the simulated annealing in economics is the work of Goffe et al. (1994) who suggested that simulated annealing could be used to optimize the objective function of various econometric estimators.
[9] A few papers consider location models where consumers are distributed on a graph. Mavronicolas et al. (2008), Feldmann et al. (2009), Nunez and Scarsini (2015), Fournier and Scarsini (2014) consider Hotelling models on graphs studying some very interesting properties of the economic system at the firm-level. See also Markey-Towler (2016) for a model that could possibly be expanded to make demand contingent upon consumers' interactions.

The rest of the paper is organized as follows. In Section 2 the agent-based model (ABM) is introduced, section 3 gives the simulation strategy and discusses the simulation results and section 4 concludes.

## 2. The market as a complex system

Following Markey-Towler and Foster (2013) and the relevant literature (Potts, 2000; Schweitzer et al., 2009; Easley and Kleinberg, 2010; Jackson, 2010), at any point in time $t$, the market is represented by a graph of a set of consumers and a set of products/firms, $G_t$, and the relationships between them represented in $l_t(G_t)$

$$E_s^t = \{G_t, l_t(G_t)\} \tag{1}$$

Where $G_t = \{g_i : g_i \in \mathbb{R}^k\}_{i=1}^{g}$ is the set of economic agents on the product-characteristics space $\mathbb{R}^k$ and $l_t(G_t) = \{l_{ij}\}_t$ is a $g \times g$ matrix that summarizes the connections between the economic agents, where, if $l_i, l_j \in E_s^t$ denotes the existence of a connection between economic agents $i$ and $j$, then

$$l_{ij} = \begin{cases} 0 & \Rightarrow g_i g_j \notin E_s^t \\ 1 & \Rightarrow g_i g_j \in E_s^t \end{cases} \tag{2}$$

In this system there are two subsets of economic agents, *consumers*, $C_t = \{c_i : c_i \in \mathbb{R}^k\}_{i=1}^{|C|_t}$, and *firms*, $F_t = \{f_i : f_i \in \mathbb{R}^k\}_{i=1}^{|F|_t}$. We consider $|F|_t$ competing firms for which $|F|_t$ different opinion networks exist. These opinion networks represent the different opinion exchange patterns of consumers for the different $|F|_t$ firms. For example, two consumers may exchange their opinions for firm A but not for firm B, hence there is a link between them apparent in opinion network A but not in network B. Suppose that in every period each consumer $i$ is represented in each network and must make a decision on buying a product from a firm, hence to stay active in one of the $|F|_t$ opinion networks. For simplicity, let us assume $f_i f_j \notin l_t(G_t), \forall f_i f_j \in G_t$, thus omitting relationships between the different opinion networks.[10] In this market, the earnings of firm $j$ immediately follow from the number of active consumers (nodes) in opinion network $j$ at the end of any given period $t$, since any active node in opinion network $j$ at the end of period $t$ represents a purchase from firm $j$.[11] Furthermore, the number of active nodes in each network is determined by the tastes and choices of each of the consumers in the market and following Simon (1955) and Tversky and Kahneman (1974) we explain below the process of these choices assuming bounded rationality.

## 2.1 Consumers' behaviour

Traditionally the assumed heuristic for consumers is a maximization rule over some utility function defined over the set of products' quantities. However, this rule is inconsistent with extensive evidence presented in behavioral economics and marketing literature. Here we assume that each consumer's preferences can be characterized by an ideal economic position in some $k$-dimensional product characteristics space, and we look closely on the decisions made by consumers when choosing which firm's product to buy. Having switched from formal analysis to computation, we depart from the classical analytically tractable models first, by assuming as baseline decision rule that consumers are affected by the opinion of their peers and second, by making the appropriate behavioral assumptions following the relevant literature of behavioral economics, discussed below.

---

[10] We assume that all consumers buy in the equilibrium (covered market is a quite standard assumption in the literature, e.g. Irmen and Thisse, 1998), where price is exogenously given. Firms do not compete on price but only on location.

[11] We assume that any product may be produced at the same constant marginal cost, which is normalized to zero without loss of generality.

Behavioral decision theory is psychological in its orientation, beginning with the view of humans as limited information processors or, perhaps more accurately, as "boundedly rational information processors" (Simon, 1955, 1956, 1957, 1959). Humans have developed a large number of cognitive mechanisms to cope with the overwhelming volume of information in the modern societies. These mechanisms are adopted automatically without any conscious and are cognitive shortcuts for making certain judgments and inferences with considerably less alternatives than those dictated by rational choice, focusing attention on a small subset of all possible information.

Kahneman and Tversky (1973, 1974, 1984) and Tversky and Kahneman (1973, 1974) have identified three general cognitive *heuristics* that decision makers adopt in the process of information gathering and analysis: (a) *decomposition,* which refers to braking a decision down into its component parts, each of which is presumably easier to evaluate than the entire decision; (b) *editing* or *pruning,* which refers to simplifying a decision by eliminating (ignoring) otherwise relevant aspects of the decision; (c) *decision heuristics,* which are simplifying the choice between alternatives thus providing cognitive efficiency.

These *heuristics* have direct application to consumers' choices. Consumers face a myriad of alternative products and there is compelling evidence which suggest that consumers simplify their decisions with a *consider-then choose* decision process in which they first identify a set of products, *the consideration set*, for further evaluation and then choose from the consideration set. In seminal observational research Payne (1976) identified that consumers use consider-then-choose decision processes. This heuristic is firmly rooted in both the experimental and prescriptive marketing literature (e.g. Bronnenberg and Vanhonacker, 1996; Brown and Wildt, 1992; DeSarbo et al., 1996; Hauser and Wernerfelt, 1990; Jedidi et al., 1996; Mehta et al., 2003; Montgomery and Svenson, 1976; Paulssen and Bagozzi, 2005; Roberts and Lattin, 1991; Shocker et al., 1991; Wu and Rangaswamy, 2003).

In the context of our model, this means that the set of alternatives each consumer, $i$, considers, $B_i^t$, is a subset of the overall set of firms/products in the system. So we can define the consideration set for consumer $i$ as

$$B_i^t = \{f_j \in F_t \setminus \|f_j - c_i\| \leq \varepsilon\} = \{f_j \in F_t \setminus z_i^t(f_j) = 1\} \subset F_t \qquad (3)$$

where $z_i^t$ is a mapping which assigns elements in the overall set of alternatives $F_t$ to either the consideration set or not, so $z_i^t : F_t \to \{0,1\}$, such that $z_i^t(f_j) = 1$ indicates that the firm/product $j$ is considered by consumer $i$ at the beginning of period $t$.

## 2.2  Consumers' networks

At the beginning of time period $t$ we consider $|F|_t$ networks, representing the consumers' different opinion exchange patterns for the $|F|_t$ competing firms. During the time period $t$, each consumer $i$ is represented in $|B|_i^t$ networks and can be active in any of these networks. In particular, $b_i^t(f_j) = 0$ if consumer $i$ is inactive in network $j$ and $b_i^t(f_j) = 1$ if consumer $i$ is active in network $j$. At the end of time period $t$, the activity of a consumer in network $j$ corresponds to the consumer's purchase of firm $j$ 's output, hence each consumer can be active only on one of the $|B|_i^t$ networks on the purchase moment (i.e., if $b_i^t(f_j) = 1$ then $b_i^t(f_m) = 0$ for $m \neq j$). Nevertheless we leave to the consumer the freedom not to make a purchase (i.e., $b_i^t(f_1) = b_i^t(f_2) = ... = b_i^t(f_{|B_i^t|}) = 0$ ). Moreover, consumers are influenced by their network peers, hence we assume that, on the purchase time, if the majority of consumer $i$'s peers are active in network $j$, the consumer will be active in the same network $j$, provided that she is not already active in another network.

Based on the above discussion and following Halu et al. (2013), the $\left|B\right|_i^t$ mathematical constraints that consumer $i$'s preferences need to satisfy at the end of period time $t$ are:

$$b_i^t(f_j) = \left[1 - \prod_{\lambda \in Nb_j^t(c_i)}(1 - b_\lambda^t(f_j))\right] \prod_{\substack{k=1 \\ k \neq j}}^{|B|_i^t}(1 - b_i^t(f_k)), \quad \text{for} \quad j = 1, 2 ... \left|B\right|_i^t \quad (4)$$

where $Nb_j^t(c_i)$ is consumer $i$'s network where she exchanges opinion with other consumers about firm $j$. Though we assume no buyer's remorse, we allow for some conflicts in the system before the purchase time, in the sense that the $\left|B\right|_i^t$ constraints provided by equations (4) will not be satisfied. The behavioral literature points out that consumers face an "uncertainty reduction" to the choice of the product they will purchase as the purchase time comes closer. The agents change and adapt their behaviour over time, reflecting both their own learning and changes in their environment. When they think of a purchase, initially they are more likely to change opinions and examine different options. They have some form of opinion at the beginning, but as they gain experience and acquire more information, their opinion changes and stabilizes over time (see Hoeffler and Ariely, 1999; West et al., 1996). This "uncertainty reduction" is captured by dynamics that slow down until the purchase time when they become completely frozen. These dynamics are implemented with the simulated annealing algorithm.

## 2.3   Evolution dynamics during consumers' purchasing process

To model how consumers decide during the time period $t$ we consider the following equation (5) that counts the number of the constraints in equations (4) that are violated (Halu et al., 2013):

$$H = \sum_{j=1}^{|B|_i^t} \left\{ \sum_{i=1}^{|C|_t} \left\{ b_i^t(f_j) - \left[ 1 - \prod_{\lambda \in Nb_j^t(c_i)} (1 - b_\lambda^t(f_j)) \right] \prod_{\substack{k=1 \\ k \neq j}}^{|B|_i^t} (1 - b_i^t(f_k)) \right\}^2 \right\} \quad (5)$$

Initially, active nodes in each network $j$ are distributed according to consumers' consideration sets [equation (3)], and we allow the consumers to change opinion at any time before they make their final choice. These dynamics are implemented using the simulated annealing algorithm, which works as follows. Starting from a relatively high initial temperature, $T$,[12] i.e. a large number of potential conflicts [a large-number of non-binding constraints given by equations (4), hence a high $H$ given by equation (5), which means that before the purchase time, consumers can be active in more than one network, i.e. they can have positive opinion for more than one firm/product] due to the stochastic way initial opinions are distributed, we use a Monte Carlo dynamics which will reach an equilibrium following the equation (5). As the time for the final choice approaches, i.e. as the simulated annealing progresses, the effective temperature $T$ decreases and the consumers tend to have less and less opinion conflicts with their network peers until they reach to zero conflicts at equilibrium, when $H = 0$.

At the end of the simulated annealing calculation we will have the configuration of the model for the period $t$, which is depicted in Figure 1. In this final configuration, each consumer can be active only on one of the $|B|_i^t$ networks or she can be inactive in all networks (no purchase). Hence, we will have "regions" in the average connectivity of consumers' opinion networks space where the active nodes of some networks will percolate the economy, while the nodes of the remaining layers will be concentrated in disconnected clusters indicating high market share inequality (since the active nodes in each network represent the respective firm's customers). Nevertheless, "regions" where the average connectivity of the opinion networks will

---

[12] Since we start with a sufficiently high temperature, the dynamics are not affected by the initial conditions of the system.

be comparable manifest a market that sustains low market share inequality (see also Halu et al., 2013).[13]

## 2.4 Firms' 'behaviour'

Following Laver and Sergenti (2011) we assume that firms' managers use an adaptive decision rule to set product characteristics on a multidimensional product-characteristics space at any given period $t$. We assume an adaptive rule that models a manager who constantly modifies product characteristics in the search for more customers, and the manager cares only about the firm's market share. If the manager's decision for the characteristics of the product at time $t$ was rewarded by an increase in market share, then the firm makes a unit move at time $t + 1$ in the same direction as the move at $t$. If not, the manager reverses direction and makes a unit move on a heading randomly selected within the half-space being faced now. In other words, the firm's manager relentlessly forages in the product-characteristics space, always searching for more customers and never being satisfied, changing strategic planning in the same direction as long as this is rewarded with more sales, but casting around for a new strategic planning when the previous one was punished with falling or static sales.

Now that we have structurally defined the evolution process of the complex system as a whole, we can proceed to simulations.

---

[13] These results arise intuitively -and are confirmed by our model- from the percolation theory on networks literature (see Newman, 2010) and the recent results concerning the role of densely connected social networks on the adoption of a behavior (see Centola, 2010).

**Figure 1: Illustration of the system's final configuration**

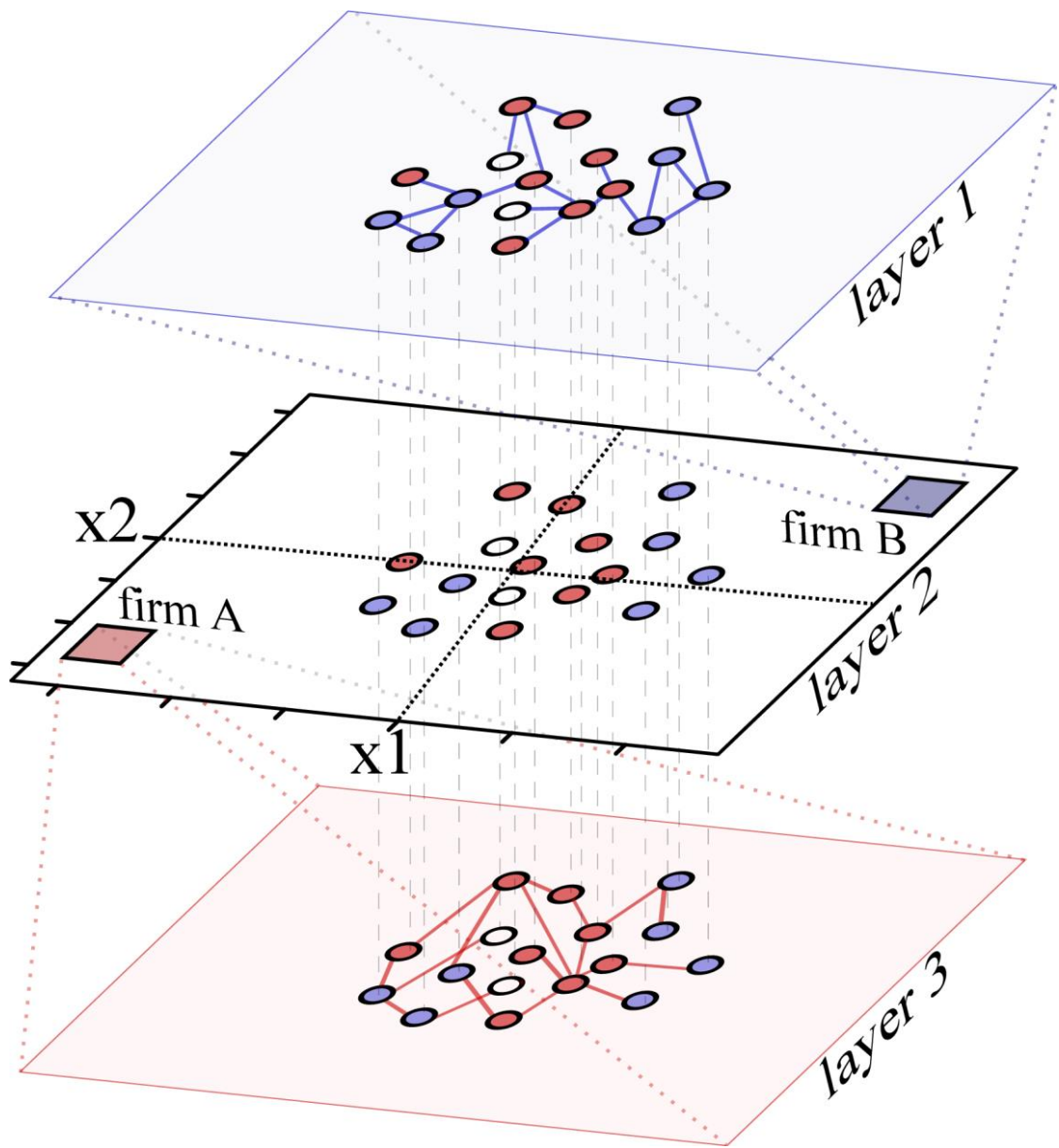

Notes: (Colored figure) *Layer 1* represents consumers' opinion network for firm *B*; *Layer 3* represents consumers' opinion network for firm *A*; *Layer 2* is the (2-dimensional) product-characteristics plane; Each consumer is represented in all layers but can be active either in the firm *B*'s underlying opinion-network (blue node) or in the firm *A*'s underlying opinion-network (red node) or inactive in both networks (white node).

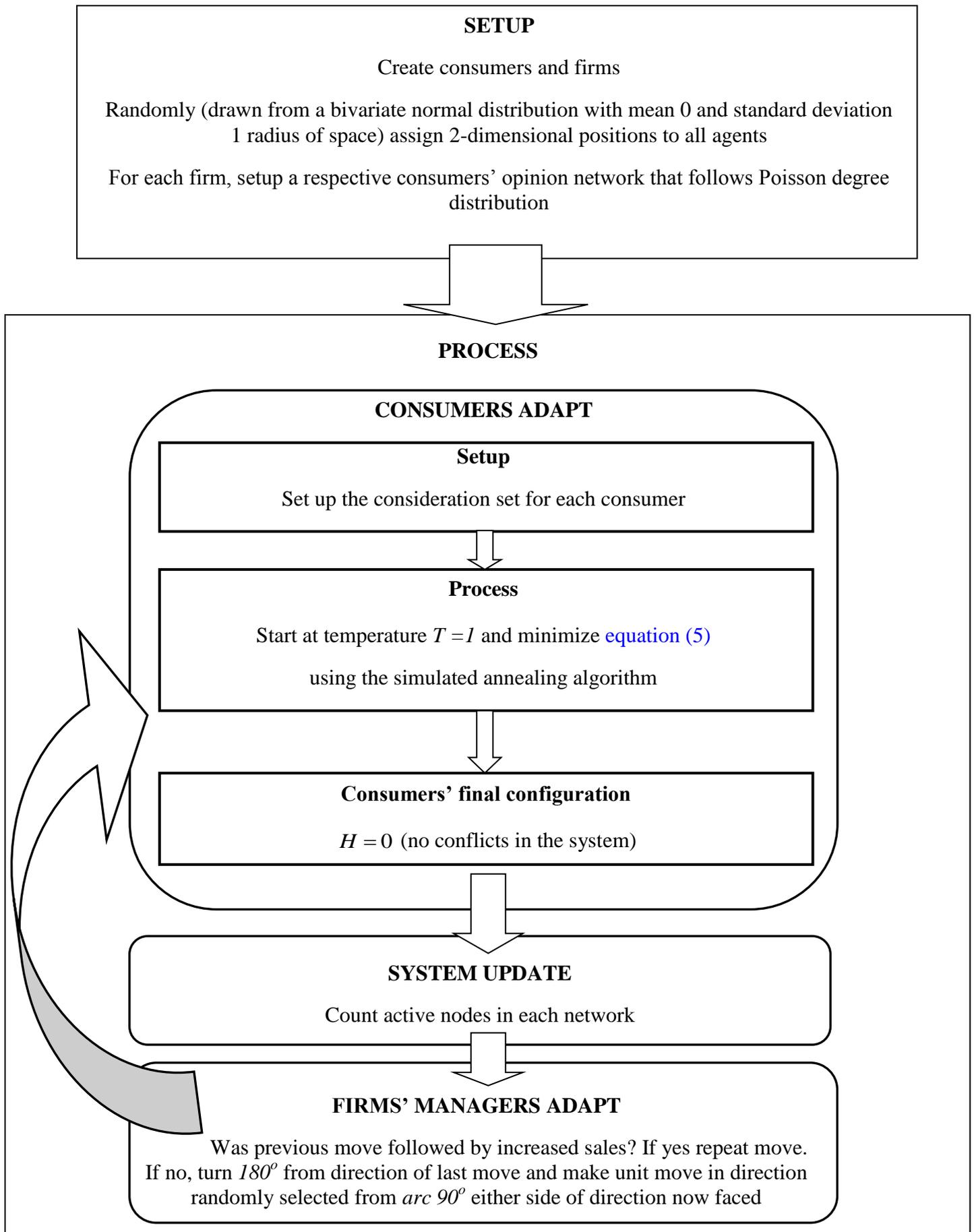

**Figure 2: Overview of the evolution of the system**

# 3. Simulations

For clarity in what follows we consider two firms, hence two layers in the consumers' opinion network, in the simulation experiments. Though, our code can be straightforwardly generalized to consider multiple firms/networks. In our setting, consumers are assumed to be intrinsically interested in product characteristics and to have ideal points in a product-characteristics space (again, in order to aid visualization, the simulated version of the model is implemented in two dimensions, i.e. in the Euclidean plane, but the model can be implemented in any number of dimensions). Firms compete with each other by offering products with varied characteristics to consumers. Figure 2 summarizes the model and its evolution process, which was programmed in R.

## 3.1 Firm System Dynamics

Initiation of the model randomly distributes a discrete set of firms' locations and consumers' ideal points across the product-characteristics plane.[14] Consumers are initially present and active in the opinion-networks/firms which lie within their consideration sets. As discussed above (see section 2.3), we allow the consumers to change opinion at any time before they make their final choice, and we model their dynamics using the simulated annealing aiming to reduce to zero the number of conflicts with their network peers (the number of violated mathematical constraints given by equation (4) for all consumers) when reaching equilibrium. More precisely, following Halu et al. (2013), the algorithm starts with an initial temperature $T = 1$ and at every time step we select a node at random from either one of the two networks with equal probability and we change it from active to inactive or vice versa. After this change, the equation (5) is recalculated, and if the difference with respect to its previous value, $\Delta H$, is negative i.e., the number of conflicts in the system is reduced, the change is accepted. If $\Delta H > 0$, the change can still be accepted but with a small probability given by $e^{-\Delta H/kT}$. This random selection process is repeated $2N$ time steps, in order to update the whole system on average once and then we advance the system time by one Monte Carlo step. The whole process is repeated by slowly reducing the temperature until we reach equilibrium at $H = 0$, where there are no more conflicts in the network. The final configuration of the "uncertainty reduction" process just described is depicted in Figure 1.

---

[14] We assign 2-dimensional positions to all agents, drawn from a bivariate random normal distribution with mean zero and standard deviation one radius of space.

We assume that consumers can share opinions with others randomly, therefore the consumers' opinion networks follow Poisson degree distributions, i.e. they fall in the class of Erdos-Renyi random networks (in the top panel of Figure 3 we illustrate how these networks look like for different average degrees). Using average degrees above the percolation threshold, we visualize the evolution of the giant component against the average degree of the network (see Newman, 2010). In the bottom panel of Figure 3 the final configuration of the two-firm model for different average-degrees of consumers' opinion networks is shown. In particular we contour-plot the difference between the total number of customers of firm A (total number of nodes active in network A) and the total number of customers of firm B (total number of nodes active in network B) as a function of the average connectivities of the two networks, $\langle k \rangle_A$ and $\langle k \rangle_B$ respectively.[15]

The simulation results visualized in Figure 3 are in accordance with the findings of Halu et al. (2013) and imply that the firm with the most connected customers gets the lion's share of the market:

- *Case 1*: In the regions where the mean degree of both networks is smaller than one, there are no giant components in the networks. This means that the networks are fragmented and the opinion of a consumer is not affected by peer opinions. In this case each firm possesses only a marginal market share and noticeable market share inequality is unlikely.
- *Case 2*: In the regions where the mean degree of either one of the two networks is smaller than one and for the other network bigger than one, the giant component in the latter network emerges and market share inequality occurs.

---

[15] For all the experiments we pin down consideration set's radius, at $\varepsilon = 0.7$. We have also simulated the model with different values of $\varepsilon$ obtaining similar results.

**Figure 3:** **The difference between the total number of firm A's customers and the total number of firm B's customers as a function of the two underlying opinion networks' average connectivities, $\langle k \rangle_A$ and $\langle k \rangle_B$.**

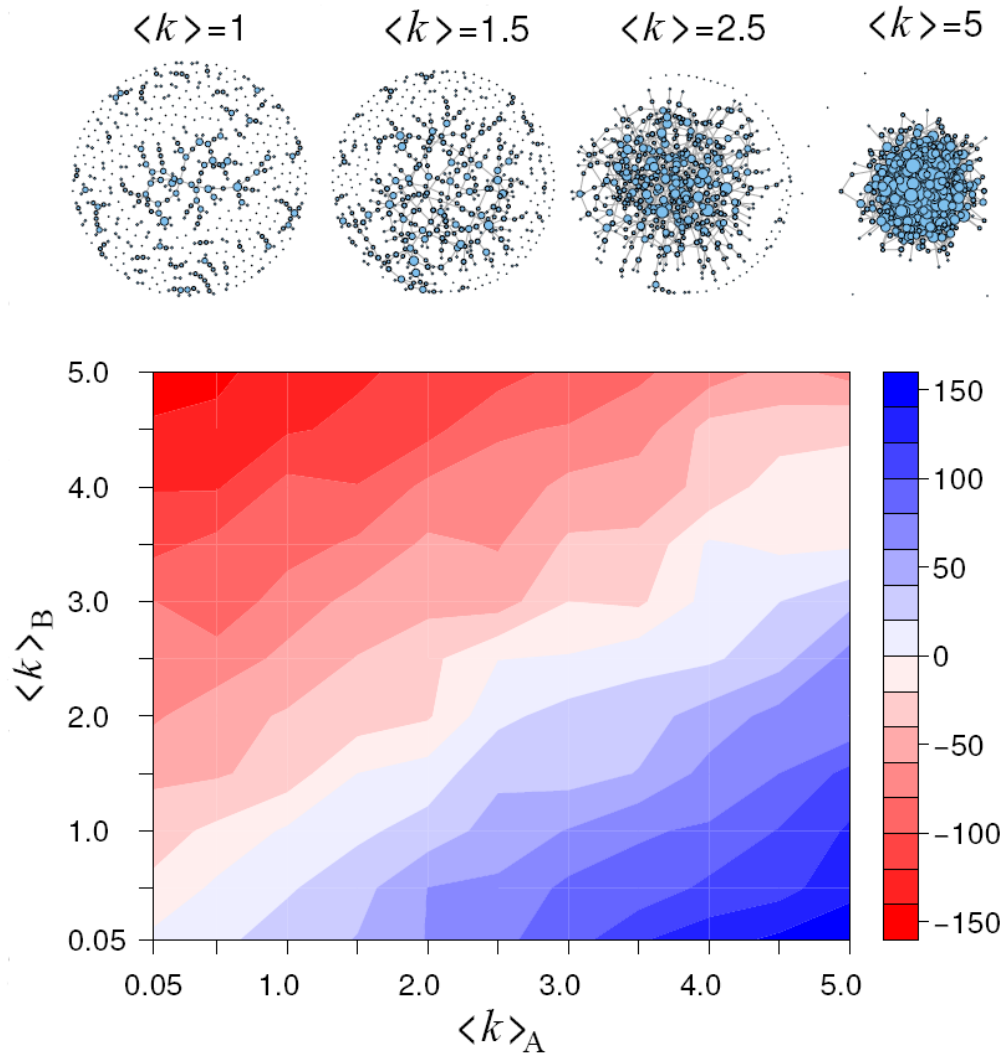

Notes: (Colored figure) *Top panel:* Examples of Erdos Renyi networks with 500 nodes and different average degrees $\langle k \rangle$. Note that while the first graph is at the percolation threshold and the largest cluster has already emerged, the graph is still very sparse. *Bottom panel*: The contour plot for the excess number of customers of firm A over firm B, $sA-sB$, for a system of 500 customers that evolved for 1000 steps. The underlying social networks are Erdos-Renyi random graphs with average degrees $\langle k \rangle$ as shown in the axis of the figure. The results are averages of 100 realizations. Consumers' consideration sets' radius is pinned down at $\varepsilon = 0.7$.

- *Case 3*: In the regions where the mean degree of both networks is greater than one, giant components emerge in both networks. In these regions we have the pluralism solution of the consumers' choices and hence, no noticeable market share inequality is apparent.

Once consumers have purchased products, firms' managers adapt their locations to reflect the pattern of consumers' preferences. Firms' managers are assumed to use an "unconstrained" adaptive rule that constantly modifies product characteristics in the search for more customers. The rule searches for customers using a "win-stay, lose-shift" algorithm (Nowak and Sigmund,1993; Bendor et al., 2003; Laver, 2005): if the previous move increased market share the manager makes another unit move in the same direction. If the previous move did not increase market share, the manager makes a unit random move in the opposite direction chosen randomly within the half space toward which it now faces.[16] Managers use no information whatsoever about the global geography of the product-characteristics plane. They have no knowledge of the ideal point of any consumer but applying recursively the limited feedback from their local environment they pick up effective clues about the best direction in which to move.

Once firms' managers have adapted firms' positions, consumers readapt and once more are active in the opinion-networks/firms which lie inside their consideration sets. Then they are again affected by peers' opinions till their final decision on which product to purchase, then firms readapt to the new configuration of consumers' preferences and the process iterates continuously.

## 3.2 Location choice

In the previous section we analyzed consumers' behaviour when their opinions are affected by the opinions of their network peers and the effect of this process in firms' market share. Next, we consider firms' location choice on the product-characteristics plane with origin (0,0) assuming that consumer ideal points' distribution is normal on

---

[16] The manager makes a unit move in a random direction on the first iteration.

**Figure 4:** The average distance $\langle d \rangle$ of the two firms ($A, B$) from the origin of the product-characteristics plane $(0,0)$, as a function of the logarithms of the underlying opinion-networks' average degrees, $\langle k \rangle_A$ and $\langle k \rangle_B$.

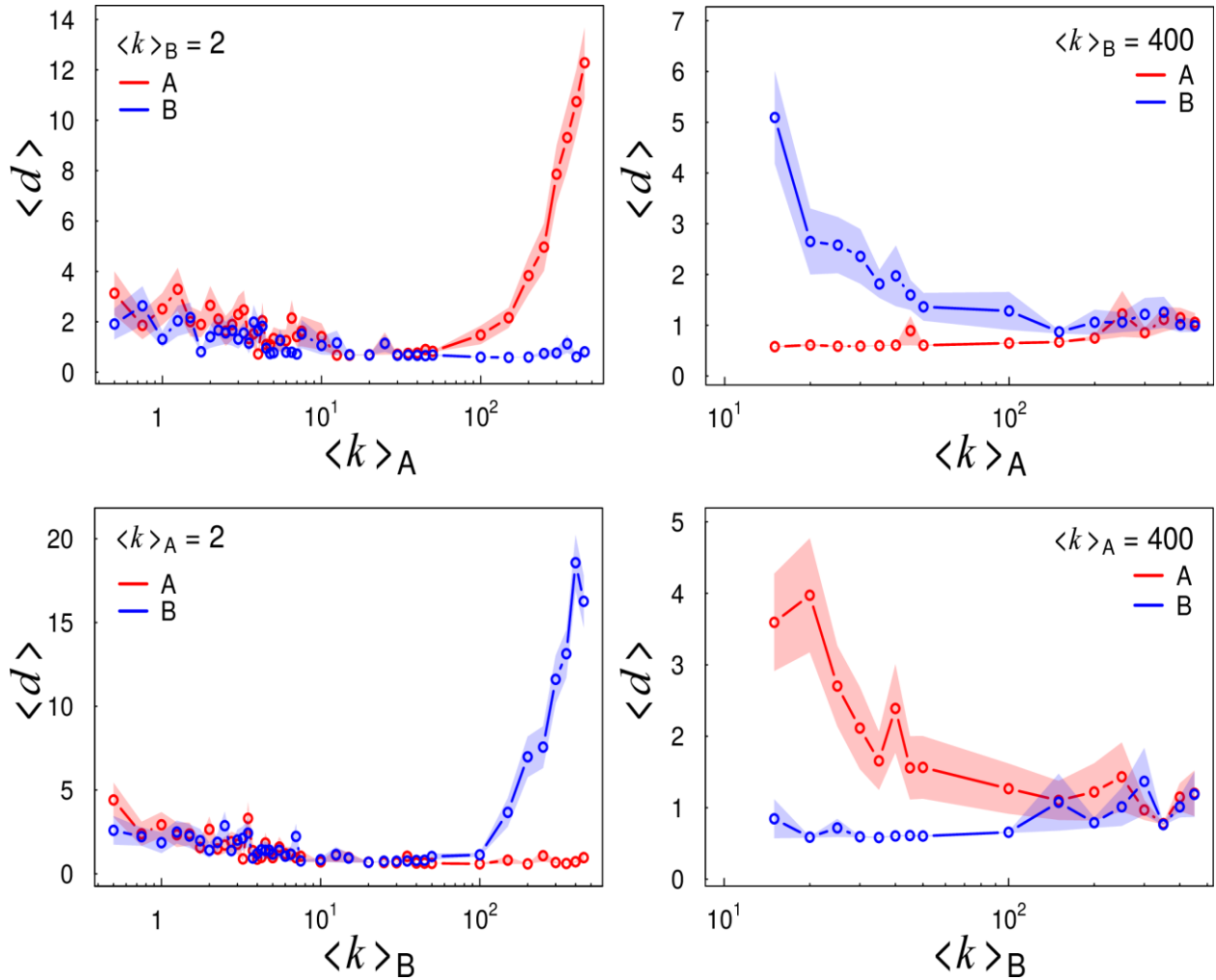

Notes: (Colored figure) The system consists of 500 consumers whose ideal points/tastes for product-characteristics are distributed following the bivariate $N(0,1)$. The distance $\langle d \rangle$ is calculated at the end of a 1000 step trajectory, and each point is averaged from 100 realizations. The shaded area around the curves represents the standard error in the calculation of the mean value. Consumers' consideration sets' radius is pinned down at $\varepsilon = 0.7$.

both product-characteristics' dimensions.[17] We check whether certain product differentiation patterns emerge.

Figure 4 summarizes the results of a simulation involving 100 independent 1000-cycle runs of the system with two firms (*A* and *B*) and 500 consumers. To avoid any dependence from the initial configuration we discard output from the first 150 cycles of each run, before the 1000 cycles were recorded. Figure 4 shows the average distance $\langle d \rangle$ of the two firms from the origin of the product-characteristics plane (0,0) calculated at the end of a 1000 step trajectory, and each point is averaged for the 100 realizations. The shaded area around the curves represents the standard error in the calculation of the mean value. For each panel we fix the average degree of the underlying opinion-network of one firm to either a very large or a very low value, and we measure $\langle d \rangle$ as a function of the average degree of the other opinion-network/firm.

From the left panels we see that, despite some fluctuations, the firms are moving closer to the center of the plane when both their underlying opinion-networks have relatively small average degrees, but when the average degree of only one of the two networks increases enough ($\langle k \rangle > 100$), the corresponding firm moves away from the center while the other firm still moves towards the center of the plane. The right panels represent the reverse behaviour; if one firm has an underlying opinion-network with large average degree and the other one with low, we find again that the one with the high degree roams anywhere in the product-characteristics plane, on average further away from the center than its competitor. On the other hand, the firm with the small average degree network stays close to the center in order to keep its consumers. However, by keeping fixed the large average degree and allowing the small average degree to increase, we observe that the firm which corresponds to the fixed degree systematically moves toward the center of the product-characteristics plane as the average degree of the network of its competitor increases, and when the average degrees are comparable, the two firms move at similar distances from the center.

Summarizing, Figure 4 discloses the following scenarios for the different phases of the model:

---

[17] Bivariate random normal distribution with mean zero and standard deviation one radius of space.

- *Scenario I:* $\langle k \rangle_A \gg \langle k \rangle_B$. In this case, firm $B$ moves toward the center of the product-characteristics plane searching for customers while the firm $A$ with better connected consumers can afford to roam anywhere in the product characteristics plane.
- *Scenario II:* $\langle k \rangle_A \approx \langle k \rangle_B$. In this case, no firm has a significant advantage over its competitor due to the presence of the opinion-network. Therefore, very much along the lines predicted by the traditional spatial competition model, both firms *systematically move toward the center of the product-characteristics plane*.
- *Scenario III:* $\langle k \rangle_A \ll \langle k \rangle_B$. In this case, it is firm $B$ that can afford to roam anywhere in the product-characteristics plane when the firm $A$ with the poorly connected underlying opinion-network is hunting customers moving towards the highest density of consumers.

Constructively, the different phases of the model show that for the case of strong inequality in the density of firms' underlying opinion-networks, the firm with the highly connected network is roaming anywhere in the product-characteristics plane away from its origin, despite the fact that this is the location of the highest density of consumers. The opposite is true, with the firms approaching the center more often, when their underlying networks are weak. The behaviour behind this pattern is easily understood if the system is watched in motion (see supplementary videos). Firms' locational moves have no effect in attracting consumers since the result of the maneuver is overshadowed by the effect of the network (the existence of links-connections between the consumers). On the other hand, the absence of opinion-networks reproduces the pattern observed in many simulations of two-party systems where the agents search for votes in the location of the highest vote densities. However, when there is no network-inequality apparent, firms go up against in attracting every single potential client, both moving toward the highest density location of consumers. In this case, the two firms go to the center of the plane (0,0) and attack each other's customer bases repelling one another from the dead center, along the lines predicted by the traditional spatial competition model.

**Figure 5**: Trajectory of the system with two firms ($A, B$) for different pairs of average degrees, $\langle k \rangle_A$ and $\langle k \rangle_B$ for the respective underlying opinion-networks.

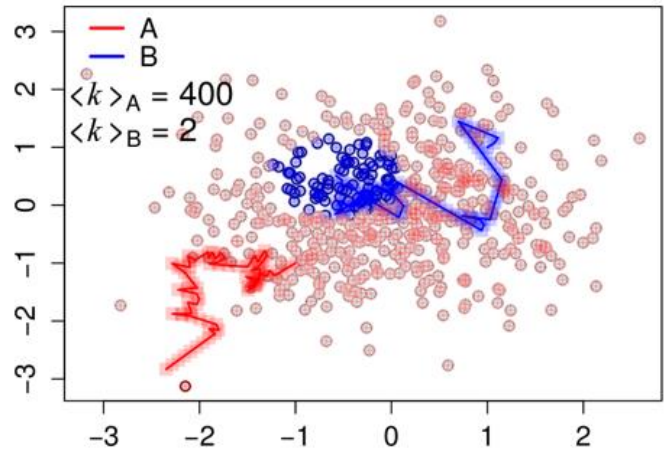

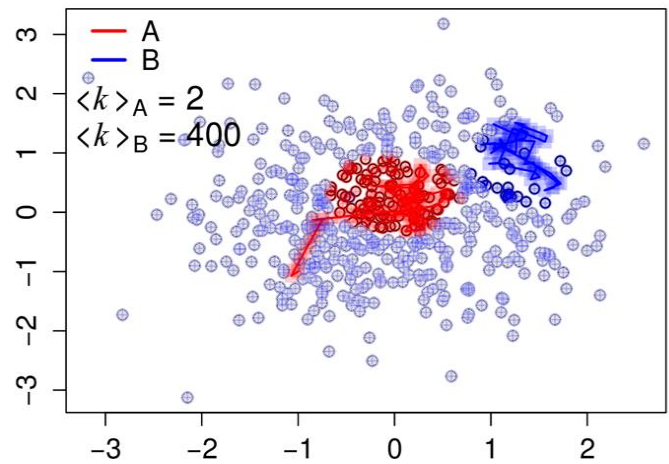

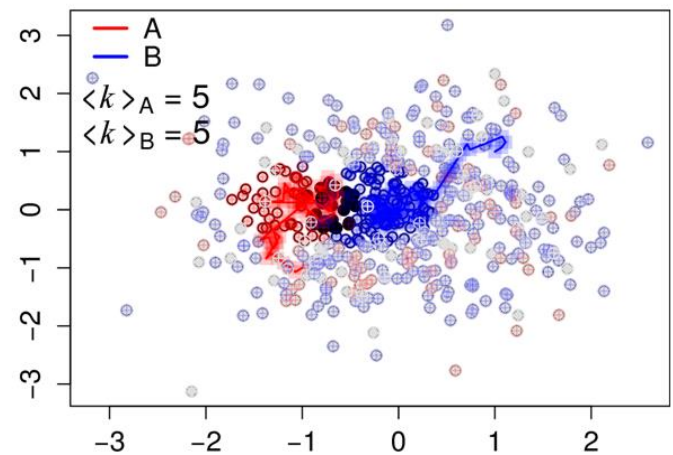

For clarity, Figure 5 shows the trajectory of the simulated system with two firms (*A* and *B*), each artificially started at the edges of the product-characteristics plane, $(-1,-1)$ and $(1,1)$, with 500 uniformly randomly scattered consumers, for different - representative for the three phases of the model- pairs of average degrees for the underlying networks: $\langle k \rangle_A = 400, \langle k \rangle_B = 2$; $\langle k \rangle_A = 2, \langle k \rangle_B = 400$; $\langle k \rangle_A = \langle k \rangle_B = 2$. The first panel shows firm *B*'s manager systematically moving toward the location of the highest consumer densities, while firm *A*'s manager enjoys the support of her customers' opinion-network and relentlessly zigzags anywhere in the product-characteristics plane in a highly atypical way (see also this video [Figure5PanelA.mp4]). The opposite pattern is apparent in the middle panel (see video [Figure5PanelB]), while the bottom panel depicts the traditional spatial model, where the two firms go to the center of the plane and attack each others' customer bases head to head (see video [Figure5PanelC]).

## 4. Conclusions

This paper contributes to the literature of product-characteristics competition at the firm-level by proposing an evolutionary economic approach that focuses on the role of consumers' social networks in the creation of new product characteristics. In the evolutionary framework, firms' competition for market share depends on the preferences/tastes of many economic agents that interact and exchange knowledge/ideas/opinions, thereby forming complex social networks which condition the evolution of the process. The influence of consumers' social networks in the development of new products is neglected in the traditional literature with 'passive' and homogeneous consumers: consumption is aggregated into economy's demand side and it is reduced to the purchase decision only, excluding social interactions. On the supply side, firms' managers incorporate new characteristics to their products that consumers can either buy or not without taking into account consumers' heterogeneous preferences and the role of their social networks.

Filling this gap, we put forth a product-characteristics competition spatial computational model based on complex systems science. Moving away from the forward-thinking strategic game theoretic models, we argue that consumers' boundedly rational behaviour within their opinion exchange networks can result in concentrated power nodes in the network structure of the market. We confirm Halu et al. (2013)'s finding that the key feature to get the higher share of the overall product-purchases is the connectivity of the consumers' opinion networks corresponding to different firms. Regarding firms' location, we use agent-based modeling to study multi-characteristics competition in the evolving market. We are interested in location competition among multiple firms in a multidimensional product-characteristics environment in which consumers and firms care for more than one product characteristic. We show that for the case of strong inequality in the density of firms' underlying opinion-networks, the firm with the highly connected network is roaming anywhere in the product-characteristics plane away from its origin, despite the fact that this is the location of the highest density of consumers. The opposite is true, with the firms approaching the center more often, when their underlying networks are weak or when all underlying networks are strong and no firm has a relative advantage on this.

Most of the relevant literature considers consumers' preferences as exogenous, referring to behavioral heuristics as determinants that lie outside the sphere of economics (Bowles, 1998). Valente (2012) argues that this assumption is used as a justification for aggregating towards a representative consumer and avoiding the economic analysis of heterogeneous agents with different preferences. Experimental and cognitive psychology empirical evidence shows that preferences are not exogenous but rather seem to be constructed during their elicitation (Shafir et al., 1993) and, furthermore, are influenced by the preferences of other agents (Tversky and Kahneman, 1981; Kahneman et al., 1982). This means that an interested party, like a firm competing for customers, might be able to shape consumers' decisions towards a specific option by influencing their preferences or affecting the opinions of their peers. Besides, it is well known that a very large share of companies' expenditures is devoted to marketing usually designed to press consumers to adopt a particular perspective/opinion of the product.

In this paper, we demonstrate how consumers' social networks can be considered the link between the supply side of the market (managers' decision on marketing and strategic directions) and the demand side (consumers' preferences influenced by managers' strategies). In this way, we capture the feedback loop between consumers and firms, namely consumers' preferences guiding firms' innovation strategies and on the other way around, firms shaping consumers' preferences by pursuing the consolidation of dense underlying social networks. The consumers' social networks are appealing to organizations not only as a low cost way of reaching an audience, but also for their increasing penetration into everyday life and decision making due to the acceleration of the social media in the web (Sadovykh et al., 2015).

Our paper is also related to the recent works by Vitell (2003), Vitell (2015), Schlaile et al. (2016), Schlaile et al. (2017), Muller (2017) on the role of consumers in responsible innovation and demand. This strand of literature illustrates that consumers' heterogeneity and bounded rationality plays an important role in the creation and diffusion of responsible innovations. The results suggest that when all consumers focus solely on negative characteristics and demand only "responsible" products and services then the complex interplay between demand and supply can create situations that are inferior to scenarios with fewer responsible consumers. Along these lines, our model could easily be expanded to include negative product characteristics allowing the identification of central actors that have the power of influencing their peers' opinions towards consuming more responsibly. Our complex networks approach may also be appropriate to address the issue of public and stakeholder engagement for stimulating responsible innovation (Jackson et al., 2005; Chilvers, 2008; Delgado et al., 2010; Owen and Goldberg, 2010; Blok et al., 2015).

Though we miss the analytical tractability of the relevant theoretical models, our simulated results demonstrate the feasibility of agent-based techniques to describe and explore firms' locational choices, as well as the capability of the proposed model to further advance the analysis in alternative ways beyond the game-theoretic framework. The proposed model should be seen only as a starting point for further research on the role of consumers in shaping markets since the real world is far more complex than our model with various aspects and factors influencing consumers' decision making.

One interesting direction for future research might be the extension of our model towards considering the role of committed agents i.e. the situation in which a fraction of consumers always remains active in one of the networks, never changing its opinion (Galam and Moscovici, 1991; Galam and Jacobs, 2007; Xie et al., 2011;

Mobilia et al., 2017; Halu et al., 2013; Masuda, 2015). Likewise, it might be also interesting to see whether the assumption of some "iconoclast consumers", i.e. consumers who intentionally behave opposite to the tendencies of their peers, alters our results.

**Funding:** Antonios Garas acknowledges funding by the EU-FET project MULTIPLEX 317532